\documentclass[prl,aps,groupedaddress]{revtex4}
\usepackage{amssymb,amsmath,amsfonts,epsfig,graphicx,latexsym,bm,color}

\begin{document}
\title{Interaction-induced chiral $p_x \pm i \,p_y$ superfluid order of bosons in an optical lattice}
\author{M. \"{O}lschl\"{a}ger$^{1}$, T. Kock$^{1}$, G. Wirth$^{1}$, A. Ewerbeck$^{1}$, C. Morais Smith$^{2}$, and A. Hemmerich$^{1}$ \footnote{e-mail: hemmerich@physnet.uni-hamburg.de} }
\affiliation{Institut f\"{u}r Laser-Physik, Universit\"{a}t Hamburg, Luruper Chaussee 149, 22761 Hamburg, Germany}
\affiliation{$^{2}$Institute for Theoretical Physics, Utrecht University, Leuvenlaan 4, 3584 CE Utrecht, The Netherlands}

\begin{abstract}
The study of superconductivity with unconventional order is complicated in condensed matter systems by their extensive complexity. Optical lattices with their exceptional precision and control allow one to emulate superfluidity avoiding many of the complications of condensed matter. A promising approach to realize unconventional superfluid order is to employ orbital degrees of freedom in higher Bloch bands. In recent work, indications were found that bosons condensed in the second band of an optical chequerboard lattice might exhibit $p_x \pm i \, p_y$ order. Here we present experiments, which provide strong evidence for the emergence of $p_x \pm i \, p_y$ order driven by the interaction in the local $p$-orbitals. We compare our observations with a multi-band Hubbard model and find excellent quantitative agreement.
\end{abstract}

\maketitle
\section{Introduction}
Understanding the role of unconventional order for superconductivity is a fundamental task in low temperature physics. Prominent examples in condensed matter are transition metal oxides \cite{Tok:00}. Studying the order parameters in these systems is complicated by their vast complexity. A widely debated example is the chiral $p_x + i \, p_y$ order possibly formed in strontium ruthenates \cite{Mae:94}, which has recently attracted much interest because its topological nature may give rise to Majorana fermions \cite{Bee:12}. Optical lattices in their lowest band have proven to be a useful experimental arena to emulate superfluidity with exceptional precision and control \cite{Lew:07, Blo:08}. However, since under most general circumstances bosonic ground state wavefunctions are necessarily positive definite \cite{Fey:72, Wu:09} and hence topologically trivial, the realization of unconventional superfluid order in optical lattices with bosons is not straight forward. Possible approaches, presently receiving great attention, are based upon amending the scalar light-shift potentials of conventional optical lattices by static abelian and even non-abelian artificial gauge fields  \cite{Lin:09, Lin:11, Dal:11, Aid:11} or upon use of dynamical lattice potentials \cite{Eck:05, Zen:09, Str:11}. A method, more closely geared to electronic matter, is to employ atoms in metastable higher bands which provide orbital degrees of freedom \cite{Isa:05,Liu:06}. Recently, we have shown that upon suitable control of band relaxation bosons can be condensed in the second band of a bipartite square lattice \cite{Wir:11}. Momentum spectra were observed consistent with chiral $p_x \pm i \,p_y$ superfluid order characterized by a spontaneously formed pattern of staggered local angular momenta, which breaks time-reversal symmetry. In the present work, by means of the following line of arguments we present clear evidence that $p_x \pm i \,p_y$ order is in fact formed: We show that for the lowest energy state of a bosonic Hubbard-model accounting for the second, third and fourth bands, the repulsive interaction in the $p$-orbitals stabilizes $p_x \pm i \,p_y$ order in analogy to Hund's second rule in multi-electronic atoms. We calculate a characteristic phase diagram with respect to a change of the interaction in the local $p$-orbitals and an adjustable distortion of the lattice, which tunes the energy minima of the second band. Experimental observations are presented, which show excellent quantitative agreement with the theoretical predictions of this phase diagram. We finally discuss excited state scenarios, which are compatible with the previously observed momentum spectra, but inconsistent with the experimental signatures reported in this work.

\begin{figure}
\includegraphics[scale=0.6, angle=0, origin=c]{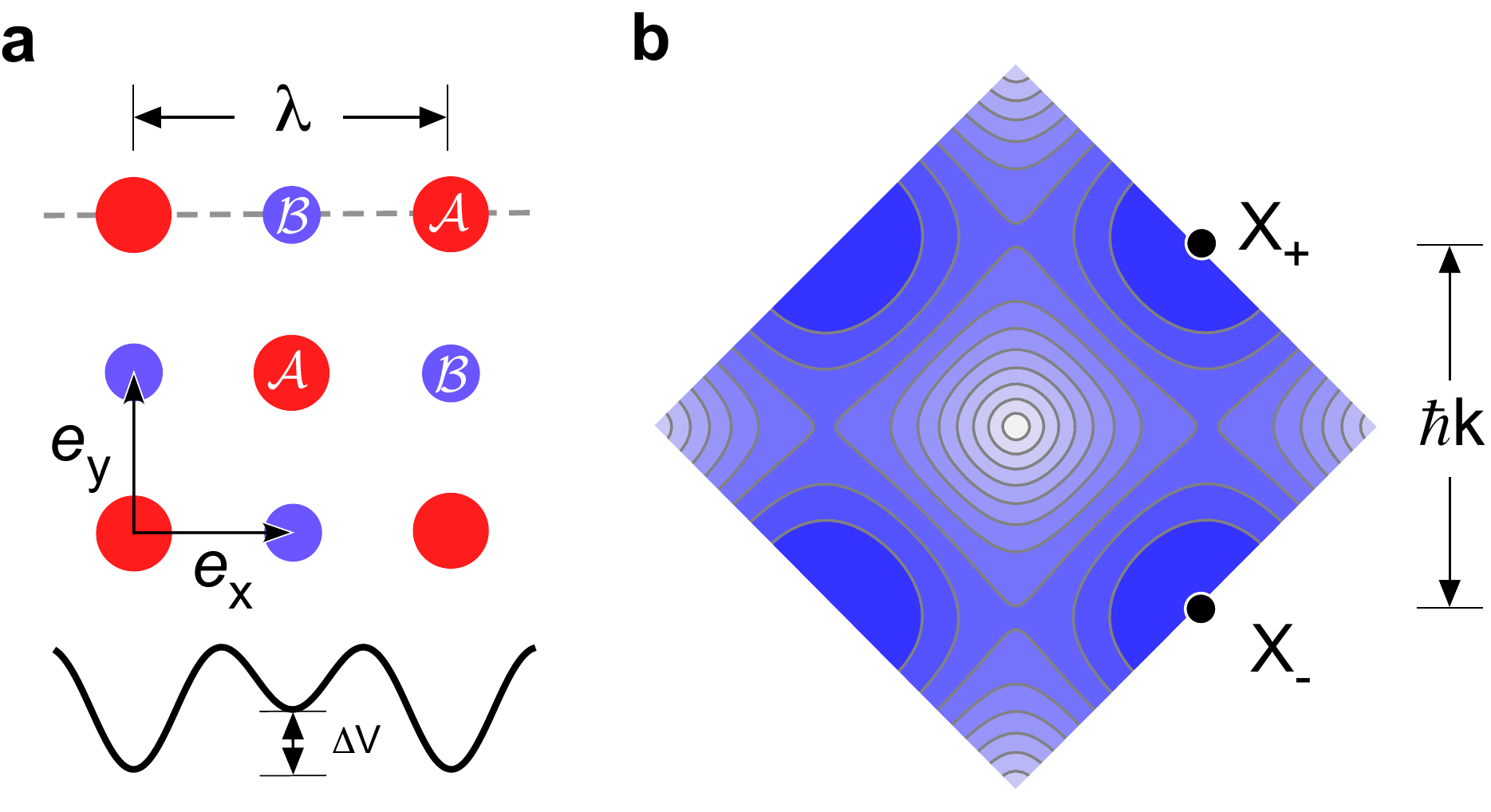}
\caption{\label{Fig.1} (a) The lattice possesses deep ($\mathcal{A}$) and shallow ($\mathcal{B}$) wells. $\lambda$ denotes the laser wavelength. A section of the potential along the dashed grey line is shown at the lower edge with the tunable relative potential energy offset $\Delta V$ indicated. (b) Contour plot of the second band within the 1st Brillouin zone for $V_0 = 6 \, E_{\textrm{rec}}$, $\Delta V = 5.7 \, E_{\textrm{rec}}$. High and low energies are indicated by light grey and dark blue, respectively. The band exhibits  two inequivalent local minima denoted as $X_{\pm} = \frac{1}{2}\hbar k\,(1,\pm1)$ ($k = \lambda/2\pi$) at the edge of the 1st Brillouin zone.}
\end{figure}

\section{Lattice potential}
We produce a two-dimensional optical potential comprising deep and shallow wells ($\mathcal{A}$ and $\mathcal{B}$ in Fig.~\ref{Fig.1} (a)) arranged as the black and white fields of a chequerboard with an average well depth $V_0$ and an adjustable relative potential energy offset $\Delta V$ \cite{Hem:91, Wir:11, Oel:11}. In the $xy$-plane the optical potential is given by
\begin{eqnarray}
\label{potential}
V(x,y) \,\equiv -\frac{V_0}{4} \,  | \, \eta \, \left(e^{i k x}  + \epsilon_{x} \,e^{-i k x} \right) + \, e^{i \beta} \left(e^{i k y} + \epsilon_{y} \, e^{-i k y} \right) |^2 \,.
\end{eqnarray}
Adjustment of $\beta$ permits controlled tuning of $\Delta V \equiv V_0 \,\eta (1+\epsilon_{x}) (1+\epsilon_{y}) \cos(\beta)$. A weak harmonic potential along the $z$-direction provides elongated tubular lattice sites. If $\eta = \epsilon_{x} = \epsilon_{y}=1$, the lattice potential possesses perfect $C_4$ rotation symmetry. In our experiment, we are constrained to fixed parameter values $\eta = 1.03$, and $\epsilon_{x} = 0.93$ and hence $C_4$ symmetry is weakly broken. In contrast to Ref.\cite{Wir:11}, careful power and polarization management of the lattice beams permits controlled adjustment of arbitrary values of $\epsilon_{y}$ around unity. The second Bloch band, shown in Fig.~\ref{Fig.1} (b), provides two inequivalent local minima at the edge of the 1st Brillouin zone (denoted by $X_{+}$ and $X_{-}$), which are energetically degenerate if the lattice displays $C_4$ rotation symmetry. By adjusting the lattice distortion parameter $\epsilon_{y}$ (see Appendix), we can continuously tune their energy difference $\Delta E \equiv E(X_{-}) - E(X_{+}) \sim 1-\epsilon_y$. In a tight-binding picture, the quantum states corresponding to $X_{\pm}$ may be approximated by Bloch states $|\psi_{\pm}\rangle$ with real-valued Bloch functions $\psi_{\pm}(r)$ composed of local $p$-orbitals ($p_x$, $p_y$) in the deep wells and local $s$-orbitals in the shallow wells. Denoting their occupations per unit cell as $n_p$ and $n_s$ with $n_0 \equiv n_s + n_p$, the relative occupations $\nu_{p} \equiv n_p/n_0$ and $\nu_{s} \equiv n_s/n_0$ only depend on the spatial shape of $\psi_{\pm}(r)$ (but not on the local chemical potential) and hence can be tuned via adjustment of $\Delta V$ (see Appendix).

\section{Multi-band Hubbard model}
In order to predict the nature of the expected quantum phases for different values of $n_{p}$ and $\Delta E$, we employ a bosonic multi-band Hubbard-Hamiltonian $\hat H$ (see Appendix) accounting for the three tight-binding bands associated with the three local orbitals $p_{x}$, $p_{y}$ and $s$. We include nearest-neighbour (NN) and next-nearest-neighbour (NNN) tunneling processes, on-site interactions for the $p$- and $s$-orbitals (as is indicated in Fig.~\ref{Fig.A1} (f) of the Appendix), and a term accounting for a potential energy difference between $p$- and $s$-orbitals, required for modeling the tuning of $\Delta V$. Furthermore, an extra term is included, which introduces a quadrupolar anisotropy of the tunneling between $p$-orbitals with respect to the $(e_{x} + \mu \,e_{y})$-directions (with $e_{x,y}$ shown in Fig.~\ref{Fig.1} (a) and $\mu \in \{-1,1\}$). This term with the real amplitude $J_{\|,a}$ permits to model the tuning of $\Delta E$ according to the relation $\Delta E \approx 8 \nu_{p} J_{\|,a}$, which is obtained by evaluating the single-particle spectrum of $\hat H$ in momentum space at $X_{\pm}$. Comparing the lowest two tight-binding bands of $\hat H$ with the second and third bands of a full two-dimensional band calculation for the experimentally realized lattice potential lets us determine the tunneling parameters. The largest tunneling amplitude $J_{sp}$ arises for NN-tunneling between $s$- and $p$-orbitals (see Appendix).

\begin{figure}
\includegraphics[scale=0.6, angle=0, origin=c]{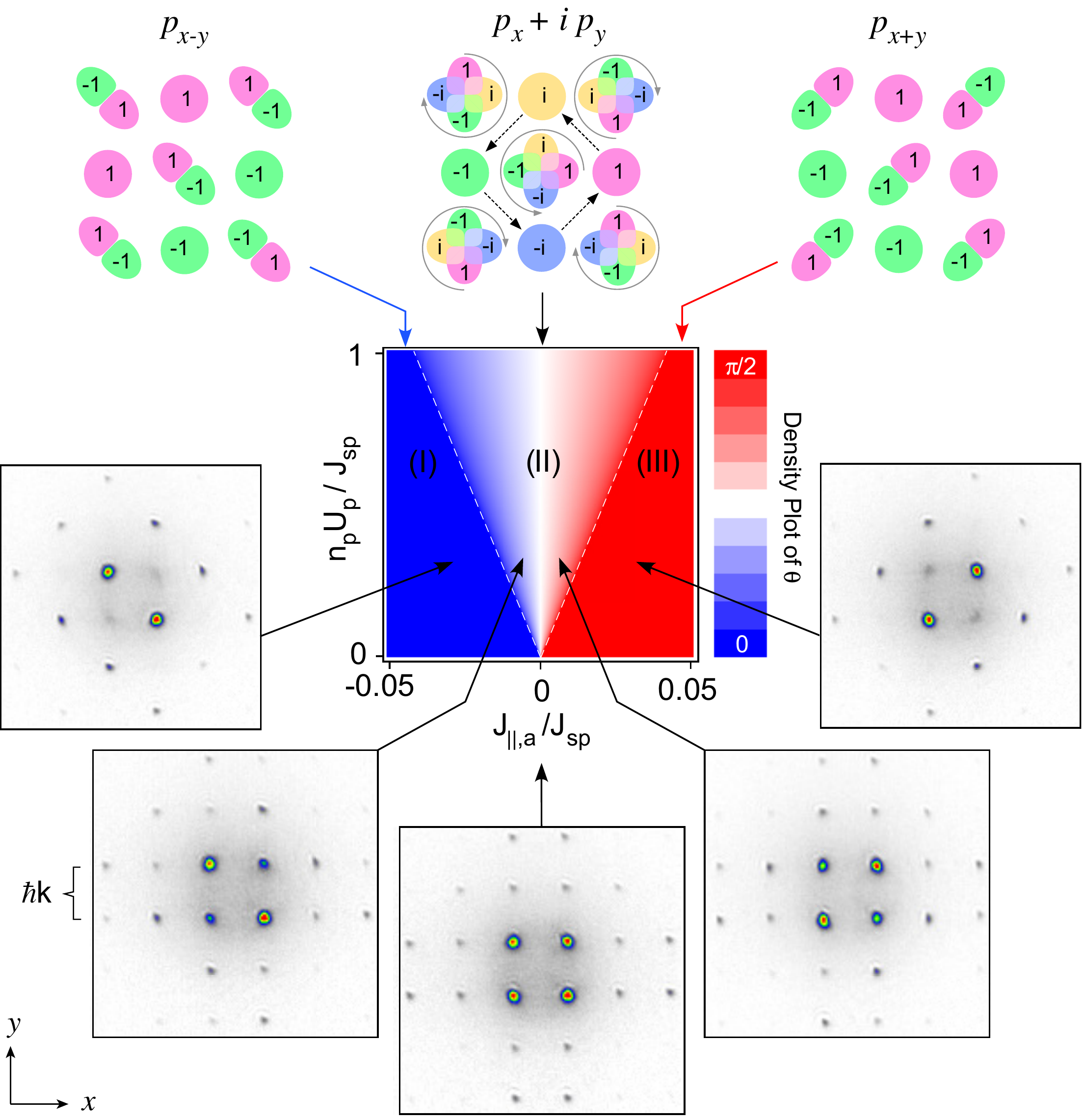}
\caption{\label{Fig.2} The phase diagram in the centre shows the mixing angle $\theta$ versus $n_{p}\,U_{p}$ and $J_{\|,a}$ (in units of $J_{sp}$). Three phases arise (regions (I),(II), and (III)) separated by 2nd order transitions (white dashed lines). On the upper edge the corresponding orbital and phase ordering is illustrated. Orbital currents and plaquette currents are highlighted by circular and straight arrows. The colors and numbers indicate the local phases of the different local orbitals. Below the phase diagram momentum spectra are shown for different values of $J_{\|,a}$.}
\end{figure}

For $\Delta E = 0$, any linear combination $|\theta,\phi\rangle \equiv \sin(\theta) \, |\psi_{+}\rangle + \cos(\theta) \, e^{i \phi} |\psi_{-}\rangle$ minimizes the single-particle energy. For repulsive collisions, minimization of the interaction energy in the $p$-orbitals requires maximal angular momentum \cite{Isa:05,Liu:06} and hence $\phi = \pm \pi/2$ and $\theta = \pi/4$ because the local superposition states $p_x \pm i p_y$ are maximally delocalized such that the atoms can optimally avoid each other. For $\Delta E \neq 0$, depending on the sign of $\Delta E$, either $|\psi_{+}\rangle$ or $|\psi_{-}\rangle$ minimizes the single-particle energy. When $\Delta E$ is increased beyond some critical value, the gain of single particle energy exceeds the cost of interaction energy required for eliminating angular momentum, and hence also the total energy is minimized by one of the states $|\psi_{\pm}\rangle$. A mean field analysis for the general case $\Delta E \neq 0$ shows that the ground state of $\hat H$ can be in fact approximated as $|\theta,\pm\pi/2\rangle = \sin(\theta) \,|\psi_{+}\rangle \pm i \cos(\theta) \, |\psi_{-}\rangle$ with the mixing angle $\theta$ plotted in the phase diagram in the centre of Fig.~\ref{Fig.2} versus $J_{\|,a}$ and $n_{p} U_{p}$ ($U_p$ is the on-site collision energy per particle in the $p_x$- and $p_y$-orbitals). Note that $\sin^2(\theta)$ and $\cos^2(\theta)$ quantify the relative populations of the condensation points $X_{+}$ and $X_{-}$, respectively. This phase diagram comprises three regions, separated by second-order phase boundaries highlighted by white dashed lines. In regions (I) and (III) one finds $\theta = 0$ and $\theta = \pi/2$ respectively (hence, only one of the condensation points $X_{\pm}$ in Fig.~\ref{Fig.1} (b) is occupied), while in region (II) the simple relation 
\begin{eqnarray}
\label{phase_diagram}
\cos(2\theta) = - \frac{12\, J_{\|,a}}{n_{p} U_{p}} = - \frac{3 \,n_0\, \Delta E}{2\, n_{p}^2\,U_{p}} 
\end{eqnarray}
holds \cite{Tie:10}. The mixing of the two condensates is governed by a competition between the gain of single-particle energy per unit cell $n_0\, \Delta E$ introduced by the lattice distortion and the interaction energy per unit cell $n_{p}^2\,U_{p}$ gained by maximizing angular momentum in the $p$-orbitals. The pictograms on the upper edge of the phase diagram illustrate the orbital and local phase ordering predicted in the three regions. In regions (I) and (III) the order parameters are real and their local phases indicated by the colors and numbers are arranged in order to maximize tunneling, while interaction energy does not play a role. In region (II) an inherently complex-valued order arises with orbital currents and plaquette currents highlighted by circular and straight arrows, respectively. Interaction energy sets the relative local phases between $p_x$- and $p_y$-orbitals at the same site to be $\pi/2$, while the phase relations between orbitals at neighboring sites are arranged to maximize tunneling. The unit cell of the order parameter comprises four unit cells of the lattice potential and time-reversal is equivalent to a shift by one unit cell of the lattice potential. Hence, time-reversal symmetry is broken. Our theoretical considerations are consistent with a numerical analysis based upon the Gross-Pitaevskii equation \cite{Cai:11} and a renormalization group analysis \cite{Liu:12}.

\section{Experiment}
Our experimental procedure begins with a Bose-Einstein condensate of rubidium atoms ($^{87}$Rb) loaded into the lowest band. By means described in Ref. \cite{Wir:11} the atoms are excited to the second band. Their temperature remains sufficiently low, such that after typically 10 ms they condense in the energy minima of the band. In our experiments we wait for 80 ms to ensure complete thermalisation. Below the phase diagram in Fig.~\ref{Fig.2}, we show momentum spectra observed in the different areas (I), (II) and (III). In regions (I) and (III) the observed Bragg-resonances unambiguously prove the realization of the predicted standing-wave order in configuration space sketched in the corresponding pictograms. In region (II), where both points $X_{+}$ and $X_{-}$ are populated, the momentum spectra appear as superpositions of those in regions (I) and (III). This clearly confirms that the underlying quantum state is composed of two states $|\psi_{\pm}\rangle$ corresponding to condensates at $X_{\pm}$ with $p_{x\pm y}$ order. However, a more precise determination of its nature requires additional information. In the following, we show that the state prepared in our experiment closely follows the phase diagram in Fig.~\ref{Fig.2} derived for the ground state $|\theta, \pm\pi/2\rangle$ of $\hat H$, which we refer to as scenario \textbf{A}.

\begin{figure}
\includegraphics[width=160mm]{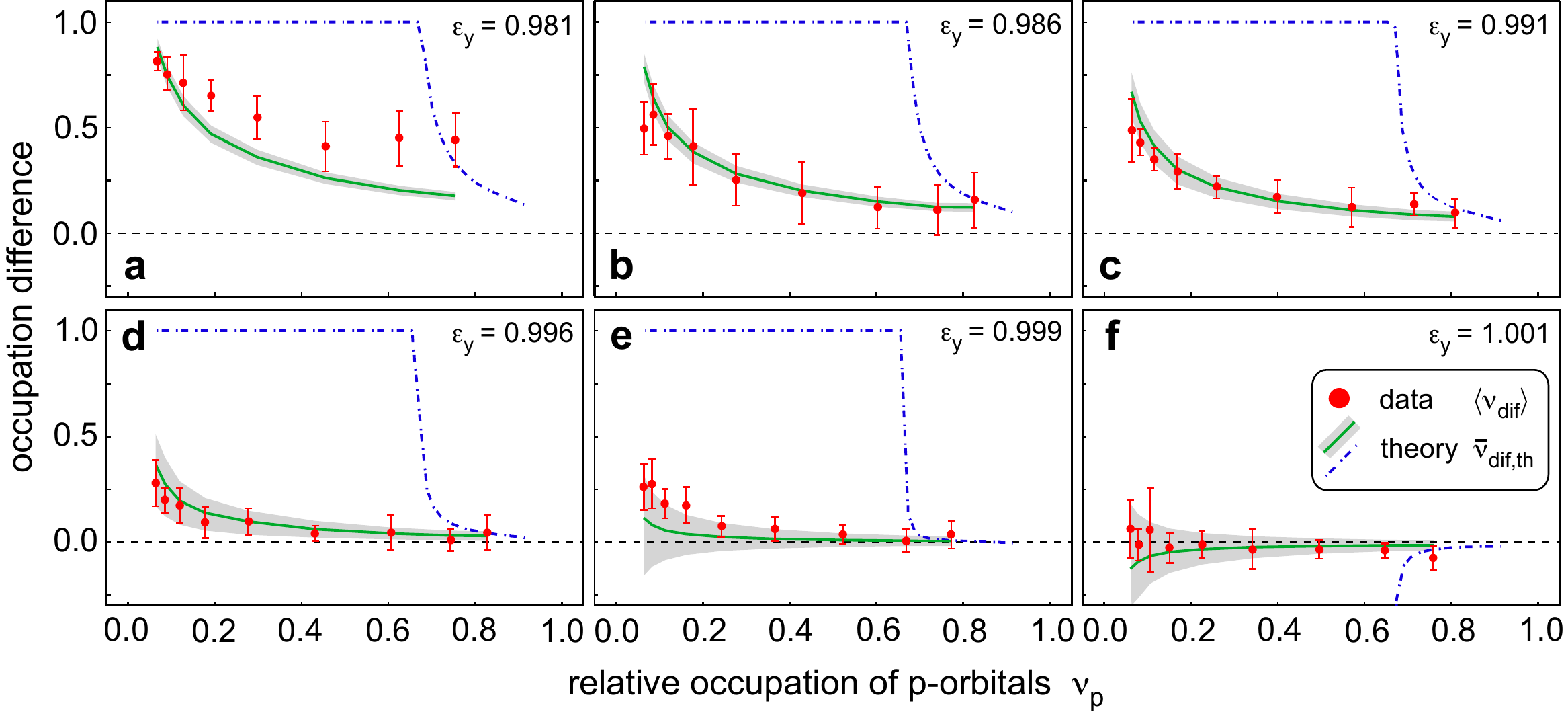}
\caption{\label{Fig.3} The normalized mean occupation difference $\langle \nu_{\text{dif}}\rangle$ between the two condensation points $X_{\pm}$ is plotted versus the relative occupation of the $p$-orbitals $\nu_{p}$. The error bars show the statistical errors for eight measurements. The adjusted distortion of the lattice potential is $1-\epsilon_y = 0.019, 0.014, 0.009, 0.004, 0.001, -0.001$ from (a) to (f). The solid green lines show the corresponding theoretical predictions $\overline{\nu}_\text{dif,th}$ derived by means of Eq.~\ref{phase_diagram}. The grey areas represent the uncertainty of the lattice distortion $\Delta\epsilon_y=\pm 2.5 \cdot 10^{-3}$. The blue dashed lines show the predictions for scenario \textbf{D}.}
\end{figure}

In Fig.~\ref{Fig.3} we compare the observed populations of the condensation points $X_{\pm}$ as a function of $\nu_p$ and the lattice distortion with the predictions of scenario \textbf{A}. The experimental procedure yielding the data points (filled red disks) in Figs.3~(a)-(f) is as follows: For different settings of the lattice distortion parameter $\epsilon_{y}$, $\Delta V$ is varied adiabatically and the normalized mean occupation difference between both condensation points $\langle \nu_{\text{dif}}\rangle \equiv (\langle n_{+} \rangle-\langle n_{-} \rangle) / (\langle n_{+} \rangle + \langle n_{-} \rangle)$ is recorded and plotted versus $\nu_p$. Here, $n_{\pm}$ are the populations observed in $X_{\pm}$ in a single measurement and $\langle \dots \rangle$ denotes the average over multiple experimental realizations. For the small interaction energies realized in our system - with $U_{p}/J_{sp}$ on the order of a few percent - and temperatures well below the condensation temperature, our mean field zero temperature analysis appears well adapted to model the observations if the finite size of the lattice with spatially varying local values of $n_p$ and $U_p$ is accounted for. In fact, upon applying a local density approximation, the observations are remarkably well reproduced by the theoretical predictions for scenario \textbf{A}. Using $\theta$ according to Eq.~\ref{phase_diagram} with the local values of $\Delta E$, $n_p$ and $U_p$, we calculate the corresponding local value of $\nu_\text{dif,th} \equiv \sin^2(\theta)-\cos^2(\theta)$ at each plaquette in the lattice and subsequently apply an average over all plaquettes to obtain $\overline{\nu}_\text{dif,th}$, which is plotted as the solid green lines in Figs.3~(a)-(f). The grey areas reflect the largest uncertainty in this calculation given by the calibration of the lattice distortion parameter. A discussion of the technical details is given in the Appendix. According to the observations in Fig.~\ref{Fig.3}, $n_+$ and $n_-$ approach each other if the distortion is reduced or if $\nu_p$ is increased, in excellent quantitative agreement with the predictions for region (II) of the phase diagram in Fig.~\ref{Fig.2}, where the value of $J_{\|,a}$ and hence $\Delta E$ required to produce imbalanced condensate fractions grows with increasing $\nu_p$. This striking behavior is hence identified as a signature of $p_{y} \pm i p_{y}$-order described by scenario \textbf{A}. The physical mechanism may be regarded as an analogue of Hund's second rule for repulsively interacting bosons: maximal local angular momentum is favorable in order to minimize interaction energy.
 
Aside from the ground state scenario \textbf{A}, the distribution of the Bragg-resonances in region (II) is also compatible with three possible excited-state scenarios: the coexistence of spatially separated real phases $|\psi_{\pm}\rangle$ (scenario \textbf{B}), a \textit{real} coherent superposition $|\theta, (1 \mp 1) \pi/2 \rangle \equiv \sin(\theta) \, |\psi_{+}\rangle \pm \cos(\theta) \, |\psi_{-}\rangle$ with $p_{x+y} \pm p_{x-y}$ order (scenario \textbf{C}), and an \textit{incoherent mixture} described by the density operator $\rho_{\theta} \equiv \sin^2(\theta) \, |\psi_{+}\rangle \langle \psi_{+}| + \cos^2(\theta) \, |\psi_{-}\rangle \langle \psi_{-}|$ (scenario \textbf{D}). The compelling quantitative agreement of our observations with the predictions for the ground state scenario \textbf{A} is underlined by the fact that the excited state scenarios \textbf{B}, \textbf{C}, and \textbf{D} lead to predictions in explicit contrast to the observations. The phase separation scenario \textbf{B} exhibits spatially separate $p_{x+y}$ and $p_{x-y}$-orbitals corresponding to the two condensates  $|\psi_{\pm}\rangle$. Hence, interaction energy does not favor equal populations of the condensation points and thus $\nu_\text{dif,th}$ cannot acquire any dependence upon $\nu_{p}$, in sharp contrast to the observations. The absence of phase separation in our experiment is also supported by the fact that in our Bragg spectra we never observe different coherence areas for the Bragg peaks corresponding to the two condensation points. This would be expected, if the measured condensate populations $n_+$ and $n_-$ would be attributed to spatially separated $|\psi_{\pm}\rangle$-domains of different size. The irrelevance of scenario \textbf{B} in our experiments is not surprising: Coexisting spatially separated domains of real phases $|\psi_{+}\rangle$ and $|\psi_{-}\rangle$ as in scenario \textbf{B} are energetically more costly than either of the pure phases $|\psi_{\pm}\rangle$. A change of the domain sizes merely requires a local redistribution of particles between the $p_{x+y}$ and $p_{x-y}$-orbitals. Particle transport over many lattice sites is not necessary. Hence, equilibration should occur within a few tunneling times leading to the formation of either $|\psi_{+}\rangle$ or $|\psi_{-}\rangle$. The real superposition state of scenario \textbf{C} exhibits spatially separate $p_{x}$ and $p_{y}$-orbitals. For equal populations of the condensation points every second local $s$-orbital remains unoccupied due to destructive interference, which is energetically strongly unfavorable. In fact, minimization of the energy of $|\theta, (1 \mp 1) \pi/2 \rangle$ with respect to $\theta$ yields either of the values $\theta=0$ or $\theta =\pi/2$ only depending on the sign of $\Delta E$ and hence $\nu_\text{dif,th}$ can only take the values $\pm 1$ (see Sec. X of Appendix). Finally, although the incoherent mixture $\rho_{\theta}$ in scenario \textbf{D} provides orthogonal $p_{x \pm y}$-orbitals at the same lattice site, the interaction energy gained by equally distributing the atoms among these orbitals is smaller than for scenario \textbf{A} because of the indeterminate phase relation between the two states $|\psi_{\pm}\rangle$. We have minimized the energy of $\rho_{\theta}$ with respect to $\theta$ in order to determine $\overline{\nu}_\text{dif,th}$ for this scenario. The result (see Appendix, Eq.A3) is plotted as the dashed blue lines in Fig.~\ref{Fig.3}, which obviously disagrees with the observations. In fact, significantly larger values of $\nu_{p}$ would be required to equilibrate the condensate fractions as compared to scenario \textbf{A}. Scenario \textbf{D} also appears implausible because the two incoherently superimposed condensates can exchange particles through binary collisions in the shared local $s$-orbitals of the shallow wells (see pictograms of regions (I) and (III)). This should rapidly degrade the coherence in each condensate, which is not observed.

\begin{figure}
\includegraphics[scale=0.6, angle=0, origin=c]{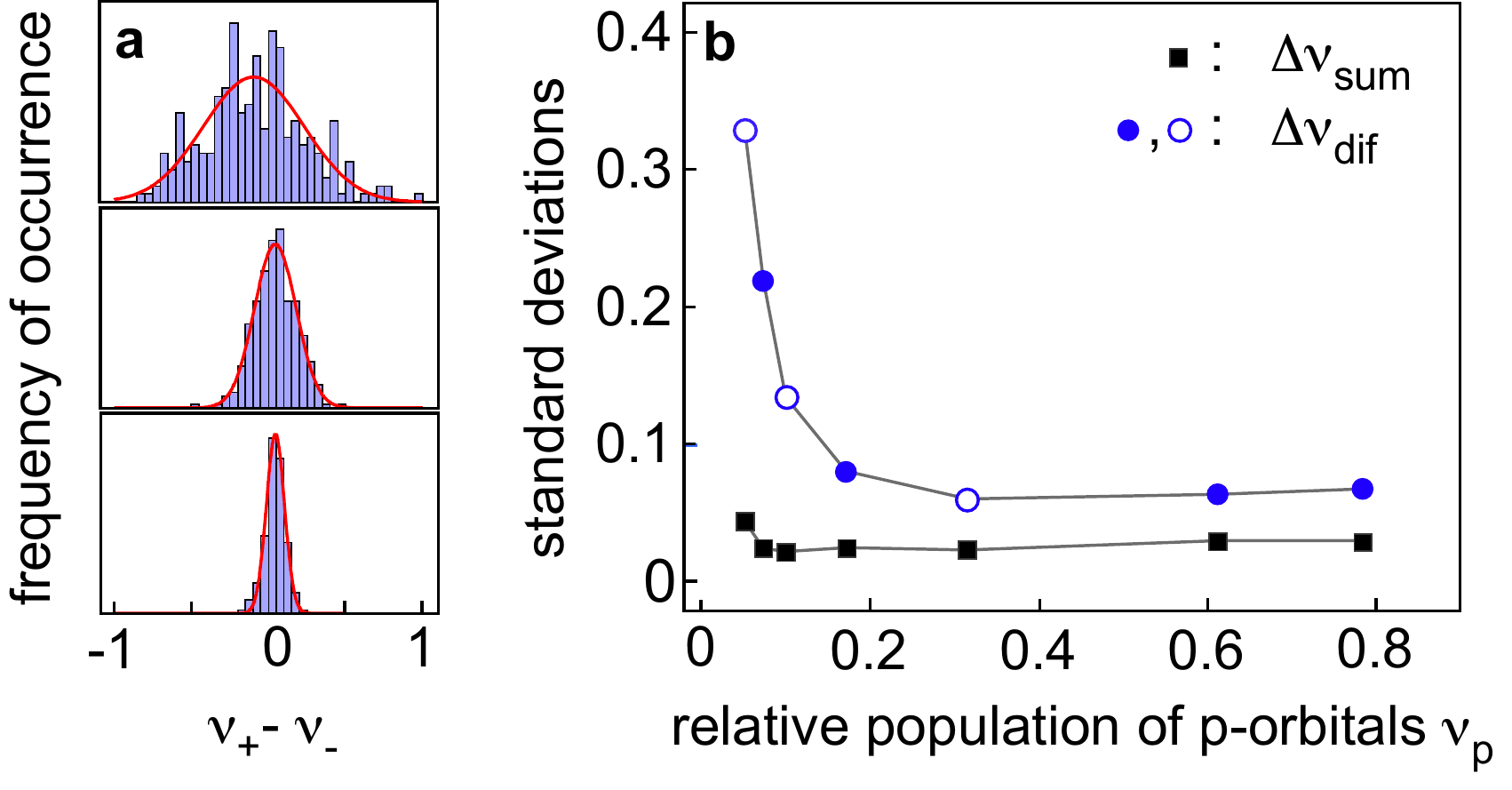}
\caption{\label{Fig.4} (a) Histograms of $\nu_{+} - \nu_{-}$ for $\nu_p$ = 0.06, 0.10, and 0.32 each showing more than 150 identical realizations. The red solid traces show Gaussian fits.  (b) The standard deviations $\Delta \nu_{\textrm{sum}}$ and $\Delta \nu_{\textrm{dif}}$ are plotted versus $\nu_p$ (open and filled blue disks: $\Delta \nu_{\textrm{dif}}$, black squares: $\Delta \nu_{\textrm{sum}}$; the open disks correspond to the histograms shown in (a); the thin grey lines connecting the data points are for guiding the eye).}
\end{figure}

Further insight is gained by analyzing the fluctuations of $\nu_{\pm} \equiv n_{\pm}/\langle n_{+} + n_{-}\rangle$. Histograms recorded for $\Delta E \approx 0$ show fluctuations of $\nu_{+} - \nu_{-}$, well described by Gaussians, centered at $\nu_{+} = \nu_{-}$ (see Fig.~\ref{Fig.4} (a)). This observation appears incompatible with the phase separation scenario \textbf{B}, for which, similarly as in Ref.\cite{Str:11}, a double peak structure should be expected instead, since equal populations $n_{\pm}$ of the two involved condensation points are not preferred. In Fig.~\ref{Fig.4} (b) the standard deviations $\Delta \nu_{\textrm{sum}} \equiv (\langle (\nu_{+} + \nu_{-})^2\rangle - \langle \nu_{+} + \nu_{-}\rangle^2)^{1/2}$ and $\Delta \nu_{\textrm{dif}} \equiv (\langle (\nu_{+} - \nu_{-})^2\rangle - \langle \nu_{+} - \nu_{-}\rangle^2)^{1/2}$are plotted versus $\nu_p$. The observations show that $\langle \nu_{+} \nu_{-}\rangle - \langle \nu_{+}\rangle \langle \nu_{-}\rangle = (\Delta \nu_{\textrm{sum}}^2 - \Delta \nu_{\textrm{dif}}^2)/4$ significantly deviates from zero, i.e., the fluctuations of $\nu_{\pm}$ are strongly correlated. While the comparably small fluctuations of the total condensate fraction $\Delta \nu_{\textrm{sum}}$ show basically no dependence on $\nu_p$, $\Delta \nu_{\textrm{dif}}$ is sizable and notably decreases as $\nu_p$ is increased, showing that the interaction in the $p$-orbitals tend to lock the populations condensed in $X_{+}$ and $X_{-}$. This observation again clearly rules out scenarios \textbf{B} and \textbf{C}, which do not prefer equal values of $n_{\pm}$ as $\nu_p$ is increased. It clearly reflects the physics of scenario \textbf{A} captured in the phase diagram in Fig.~\ref{Fig.2}. As $\nu_p$ is decreased, the critical point is approached along the vertical $J_{\|,a}=0\,$-line where all three phases adjoin. The larger $\nu_p$  the higher the interaction energy to be paid for deviations of $\nu_{+} - \nu_{-}$ from zero yielding suppression of fluctuations $\Delta \nu_{\textrm{dif}}$.

\section{Conclusion}
Optical lattices with $p_x \pm i \,p_y$ order open exciting perspectives for future research. Matter wave interference techniques \cite{Cai:12} could be used to further study the mutual coherence of the condensates at $X_{+}$ and $X_{-}$. Imaging of the atoms with single-site resolution as demonstrated in Refs. \cite{She:10, Gre:10} might allow one to directly observe the local angular momentum of the wave function and hence to explore the spontaneous symmetry breaking process. Proceeding to deeper potential wells, one may access the strongly correlated regime where a rich phase diagram of Mott insulators with distinct orbital ordering is expected \cite{Li:11, Pin:12, Mar:12}. One may also explore topologically protected features in higher bands \cite{Oel:12} and, if fermions are used, simulate forms of topological matter \cite{Sun:12} resembling those discussed in the context of electronic systems.

\section{Appendix}
\textbf{A I. Lattice potential}. Using an interferometric lattice set-up \cite{Hem:91}, we produce a two-dimensional optical potential comprising deep and shallow wells ($\mathcal{A}$ and $\mathcal{B}$ in Fig.~\ref{Fig.1} (a) of the main text) arranged as the black and white fields of a chequerboard with an average well depth $V_0$ and an adjustable relative potential energy offset \cite{Wir:11}. In the $xy$-plane the optical potential is given by
\begin{eqnarray}
\label{potential_A}
V(x,y) \,\equiv -\frac{V_0}{4} \,  | \, \eta \, \left(e^{i k x}  + \epsilon_{x} \,e^{-i k x} \right) + \, e^{i \beta} \left(e^{i k y} + \epsilon_{y} \, e^{-i k y} \right) |^2 \,.
\end{eqnarray}
Adjustment of $\beta$ with a precision exceeding $\pi/300$ permits controlled tuning of $\Delta V \equiv V_0 \,\eta (1+\epsilon_{x}) (1+\epsilon_{y}) \cos(\beta)$. A weak harmonic potential (with 40 Hz vibrational frequency) along the $z$-direction provides elongated tubular lattice sites. If $\eta = \epsilon_{x} = \epsilon_{y}=1$, the lattice potential possesses perfect $C_4$ rotation symmetry. In our experiment, due to unavoidable imperfections of the lattice set-up, we are constrained to fixed parameter values $\eta = 1.03$, and $\epsilon_{x} = 0.93$ and hence $C_4$ symmetry is weakly broken. In contrast to Ref.\cite{Wir:11}, the optical set-up permits controlled adjustment of arbitrary values of $\epsilon_{y}$ within an interval including $\epsilon_{y}=1$. This is accomplished as follows: the optical standing wave along the $y$-axis is obtained by a retro-reflected laser beam. The linear polarization of the incoming beam can be rotated with a retardation plate. After retro-reflection the polarization is rotated to precisely match with the $z$-direction, which exclusively contributes to the lattice potential. 
\begin{figure}
\includegraphics[scale=0.4, angle=0, origin=c]{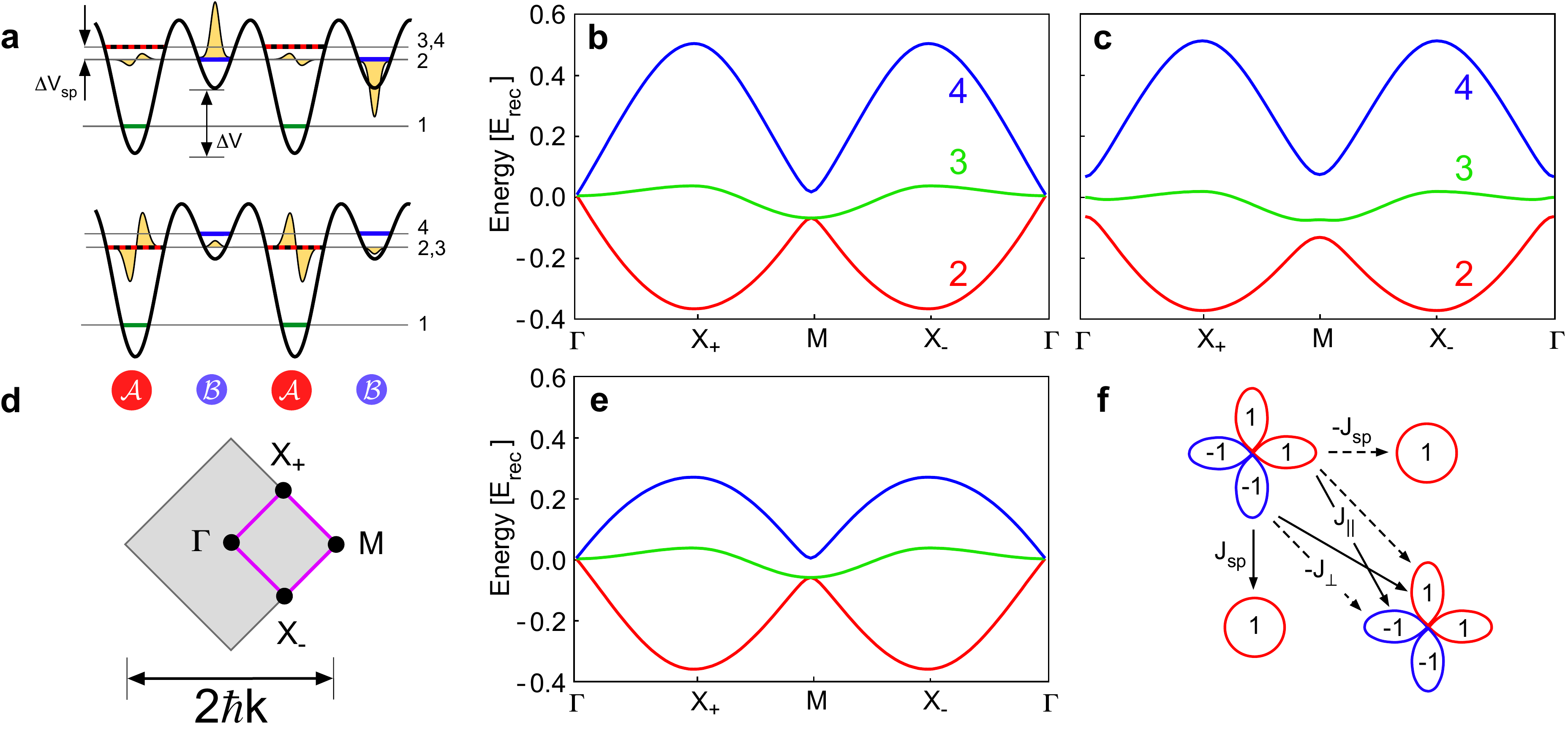}
\caption{\label{Fig.A1} (a) Schematic of first four bands plotted in configuration space for $\Delta V_{sp} < 0$ (upper detail) and $\Delta V_{sp} > 0$ (lower detail). The black and red dashed lines represent degenerate local $p_x$- and $p_y$-orbitals yielding two closely spaced bands. In (b) and (c) the 2nd (red), 3rd (green) and 4th (blue) Bloch bands are plotted along a roundtrip in the 1st Brillouin zone (grey square in  (d)) connecting the points $\Gamma, X_{+}, M, X_{-}, \Gamma$. (b) and (c) show the idealized $C_4$-symmetric case ($\eta = \epsilon_{x} = \epsilon_{y}=1$) and the experimentally implemented case ($\eta = 1.03, \epsilon_{x} = 0.93, \epsilon_{y} = 1$), respectively. (e) TB bands based upon the Hamiltonian in Eq.~(\ref{BH_Hamiltonian}) fitted to match the bands in (b). (f) Tunneling amplitudes used in the Hubbard model in Eq.~(\ref{BH_Hamiltonian}). The numbers $\nu$ plotted inside the local $s$- and $p$-orbitals denote their local phases $e^{i(\nu-1)\pi/2}$.}
\end{figure} 
\\ \\
\textbf{A II. Band structure: tight-binding picture and full-band calculation}.
One may understand the single-particle bands within a simplified tight-binding (TB) picture. As sketched in Fig.~\ref{Fig.A1} (a), each local vibrational orbital of the two types of wells gives rise to a Bloch band. We operate in the regime where the 2nd, 3rd and 4th bands are significantly closer to each other than their separation from the 1st band, which correspond to the local $s$-ground states in the deep wells. We may write $\Delta V = V_{sp} + \Delta V_{sp}$ with a potential energy offset $V_{sp}$ depending on the adjusted value of $V_0$, and $\Delta V_{sp}$ denoting the potential energy difference between the local $s$-states in the shallow wells and the local $p_x$ and $p_y$ orbitals in the deep wells (cf. Fig.~\ref{Fig.A1} (a)). For $V_0 = 6 \, E_{\textrm{rec}}$ used in our experiment, $V_{sp} = 5.7 \, E_{\textrm{rec}}$. If $\Delta V_{sp} < 0$, the second band arises primarily from the local $s$-states in the shallow wells (upper graph in (a)), while for $\Delta V_{sp} > 0$ it corresponds to a superposition of the degenerate $p_x$- and $p_y$-orbitals in the deep wells (lower graph in \textbf{a}). A two-dimensional band calculation for the potential in Eq.~(\ref{potential_A}) yields the true 2nd, 3rd and 4th Bloch bands. In Fig.~\ref{Fig.A1} (b) and (c), these bands are plotted along a roundtrip in the 1st Brillouin zone connecting the points $\Gamma, X_{+}, M, X_{-}, \Gamma$ (indicated in (d)). The average well depth is $V_0 = 6 \, E_{\textrm{rec}}$ and $\Delta V = 5.7 \, E_{\textrm{rec}}$, which corresponds to $\Delta V_{sp} \approx 0$ in the TB picture in (a). In (b) the idealized $C_4$ symmetric case is shown, compared in (c) to the experimentally implemented case with $\epsilon_{y}=1$. The band degeneracies arising at the $\Gamma$- and $M$-points in (b) are lifted in (c) due to the lattice distortion. In either case the 2nd band possesses two inequivalent band minima at the $X_{-}$- and $X_{+}$-points with energies $E(X_{\pm})$, which are degenerate even in the case (c) despite the broken $C_4$-symmetry. Only if $\epsilon_{y}$ is tuned away from unity, this degeneracy is lifted. Experimentally, tuning of $\Delta E \equiv E(X_{-}) - E(X_{+})$ is accomplished by tuning of $\epsilon_{y}$ with $\Delta E(\epsilon_{y}) \sim 1-\epsilon_y$ quantified by a band calculation for the potential in Eq.~(\ref{potential_A}). In the vicinity of the $X_{-}$- and $X_{+}$-points, where the condensed atoms reside, the bands in (b) and (c) are approximately equal.
\\ \\
\textbf{A III. Bosonic multi-band Hubbard model}. We employ the multi-band Hubbard-Hamiltonian  
\begin{eqnarray}
\label{BH_Hamiltonian}
\hat H &\equiv& 
-J_{ss} \sum_{R \in \mathcal{B}, \mu, \nu} \hat s^{\dagger}_{R}  \hat s_{R+\mu d_{\nu}}
-J_{sp} \sum_{R \in \mathcal{A}, \sigma, \mu} \bigl(  \mu \, \hat p^{\dagger}_{\sigma,R}  \hat s_{R+\mu e_{\sigma}} \, \, +   \, \, h.c. \bigr)
\nonumber \\
&-& \sum_{R \in \mathcal{A}, \sigma, \mu, \nu} (J_{\|} + \nu J_{\|,a}) \, \hat p^{\dagger}_{\sigma,R}  \, \hat p_{\sigma,R+\mu d_{\nu}}
- J_{\bot} \sum_{R \in \mathcal{A}, \mu, \nu} \nu \, (\,\hat p^{\dagger}_{x,R} \,\hat p_{y,R+\mu d_{\nu}} + \hat p^{\dagger}_{y,R}  \, \hat p_{x,R+\mu d_{\nu}})  \\
&+& \frac{\Delta V_{sp}}{2}  \sum_{R \in \mathcal{A}} \, (\hat n_{s,R+e_{x}} - \hat n_{p,R})
+  \frac{U_p}{2} \sum_{R \in \mathcal{A}} ( \hat n^2_{p,R} -  \hat L^2_{p,R}/3)  +  \frac{U_s}{2} \sum_{R \in \mathcal{B}}  \hat n_{s,R}( \hat n_{s,R}-1) \, ,
\nonumber \\ \nonumber
\end{eqnarray}
with the summation indices $\mu, \nu \in \{-1,1\}, \sigma \in \{x,y\}$, $d_{\nu} \equiv e_{x} + \nu e_{y}$, and $e_{x,y}$ as shown in Fig.~\ref{Fig.1} (a) of the main text. This Hamiltonian accounts for all possible tunneling processes between nearest- (NN) and next-nearest neighbors (NNN), as is indicated in Fig.~\ref{Fig.A1} (f): NN-tunneling between $s$-orbitals and $p$-orbitals ($J_{sp}$), NNN-tunneling between $s$-orbitals ($J_{ss}$), NNN-tunneling between $p_x$-orbitals or $p_y$-orbitals ($J_{\|}$), NNN-tunneling between $p_x$- and $p_y$-orbitals ($J_{\bot}$), a term accounting for a potential energy difference between $p$- and $s$-orbitals ($\Delta V_{sp}$) and the on-site interactions for $p$-orbitals ($U_p$) and $s$-orbitals ($U_s$), respectively, with $ \hat n_{s,R} \equiv  \hat s^{\dagger}_{R}  \hat s_{R}$, $ \hat n_{p,R} \equiv  \hat p^{\dagger}_{x,R} \,\hat p_{x,R} + \hat p^{\dagger}_{y,R}  \, \hat p_{y,R}$ and $ \hat L_{p,R} \equiv i ( \hat p^{\dagger}_{x,R}  \,\hat p_{y,R} -  \hat p^{\dagger}_{y,R} \,\hat p_{x,R})$. Furthermore, an extra term scaling with $J_{\|,a}$ is included, which introduces a quadrupolar anisotropy of the tunneling between $p$-orbitals with respect to the $d_{\nu}$-directions. This term permits to model the tuning of the energy difference $\Delta E$ between the minima of the second band. Diagonalizing the kinetic part of $\hat H$ in momentum space at $X_{\pm}$ yields the relation $\Delta E \approx 8 \nu_{p} J_{\|,a}$ with the relative occupation $\nu_{p}$ of the $p$-orbitals given as $\nu_{p}\approx \frac{1}{2} [1+ \Delta V_{sp} /(32 J_{sp}^2 + \Delta V_{sp}^2)^{1/2}]$.
\\ \\
\textbf{A IV. Determination of hopping parameters and phase diagram}. 
The Hubbard Hamiltonian of Eq.~(\ref{BH_Hamiltonian}) yields three TB bands associated with the three local orbitals involved. Setting $J_{\|,a} = 0$ we match the lowest two TB bands with the full bands in Fig.~\ref{Fig.A1} (b). This lets us determine the tunneling parameters as $J_{ss} \approx 0$, $J_{sp} = 0.12 \,E_{\textrm{rec}}$, $J_{\|} = 0.07 \, J_{sp}$, $J_{\bot} = 0.15 \, J_{sp}$, and $\Delta V_{sp} = 0.3\,J_{sp}$. The resulting TB bands are plotted in Fig.~\ref{Fig.A1} (e) showing good agreement with the full bands. The smaller bandwidth of the 3rd TB band as compared to the corresponding 4th full band indicates the proximity of higher bands ($p$-orbitals in the shallow wells) in the true lattice potential, not accounted for in the TB-description. A non-zero value of $J_{\|,a}$ introduces an imbalance for NNN-tunneling between $p$-orbitals along the $x+y$ and $x-y$-directions, which acts to lift the degeneracy of the two band minima of the lowest TB band. The phase diagram in Eq.~(1) of the main text is derived as follows: the Hamiltonian in Eq.~(\ref{BH_Hamiltonian}) including $J_{\|,a}$ is rewritten, using a composition of the lattice via plaquettes with four sites. We diagonalize the kinetic energy in momentum space and assume that the system is condensed at the two inequivalent energy minima of the lowest TB band. Replacing momentum space operators by their mean values, we calculate the interaction energy to zeroth order in the fluctuations and minimize the total energy with respect to the relative phase between the two possible condensation states and with respect to their relative weights.
\\ \\
\textbf{A V. Momentum spectra and band populations.} Momentum spectra are obtained by rapidly ($< 1\,\mu$s) extinguishing the lattice potential, permitting a free expansion of the atomic sample during 30 ms, and subsequently recording an absorption image. Band populations are measured as follows: the population of the n-th band is transferred into the n-th Brillouin zone by adiabatically terminating the lattice potential in 400 $\mu$s, followed by a ballistic expansion of 30 ms. An absorption image of the atomic density distribution is recorded and the populations in the different Brillouin zones are counted. 
\\ \\
\textbf{A VI. Tuning of $\nu_{p}$, effect of band relaxation.} The required tuning of the relative population of the $p$-orbitals  $\nu_{p}$ is accomplished via adjustment of $\Delta V$. Since this tuning is essential, we have studied it in some detail. The value of $\nu_{p}$ corresponding to some $\beta$ and $\epsilon_{y}$ (and hence $\Delta V$ and $\Delta E$) is obtained by integrating $|\psi_{\pm}(r)|^2$ over those fields of the chequerboard lattice comprising the deep wells with the Bloch functions $\psi_{\pm}(r)$ at the condensation points $X_{\pm}$ derived from a band calculation employing the potential of Eq.~(\ref{potential_A}). $\nu_{p}$ scales monotonously with $\Delta V$ (as shown by the solid black trace of Fig.~\ref{Fig.A2} (a)) and it is practically independent of the lattice distortion parameter $\epsilon_{y}$ and hence of $\Delta E$. This behavior is consistent with the approximate analytic expression derived from the Hubbard model in Sec. A III. 

Changing $\Delta V$ also has a significant impact on the time-scale of band relaxation, as is shown by the data points in Fig.~\ref{Fig.A2} (a). The band lifetime is found to be maximal ($\approx 230\,$ms) if most atoms reside in the local $s$-orbitals ($\nu_{p} \approx 0$). This is expected, since for these atoms there is no local state with lower energy available, which could give rise to relaxation. The initial preparation of the condensate is carried out in this configuration. Following a holding time of 80 ms to reach complete equilibrium, a subsequent adiabatic increase of $\nu_{p}$ during 3 ms is directly visible in the momentum spectra: the higher order Bragg-peaks become more populated due to the increased contribution of the local $p$-orbitals, which due to their nodal structure comprise higher momenta. This is shown in Fig.~\ref{Fig.A2} (b) and (c) with the corresponding calculation in Fig.~\ref{Fig.A2} (d) yielding good agreement. Increasing values of $\nu_{p}$ are accompanied by faster band decay. Since a single tunneling process between adjacent wells is sufficient for adjusting $\nu_{p}$, adiabaticity only requires tuning times of a few tunneling times of about 1 ms.

\begin{figure}
\includegraphics[scale=0.6, angle=0, origin=c]{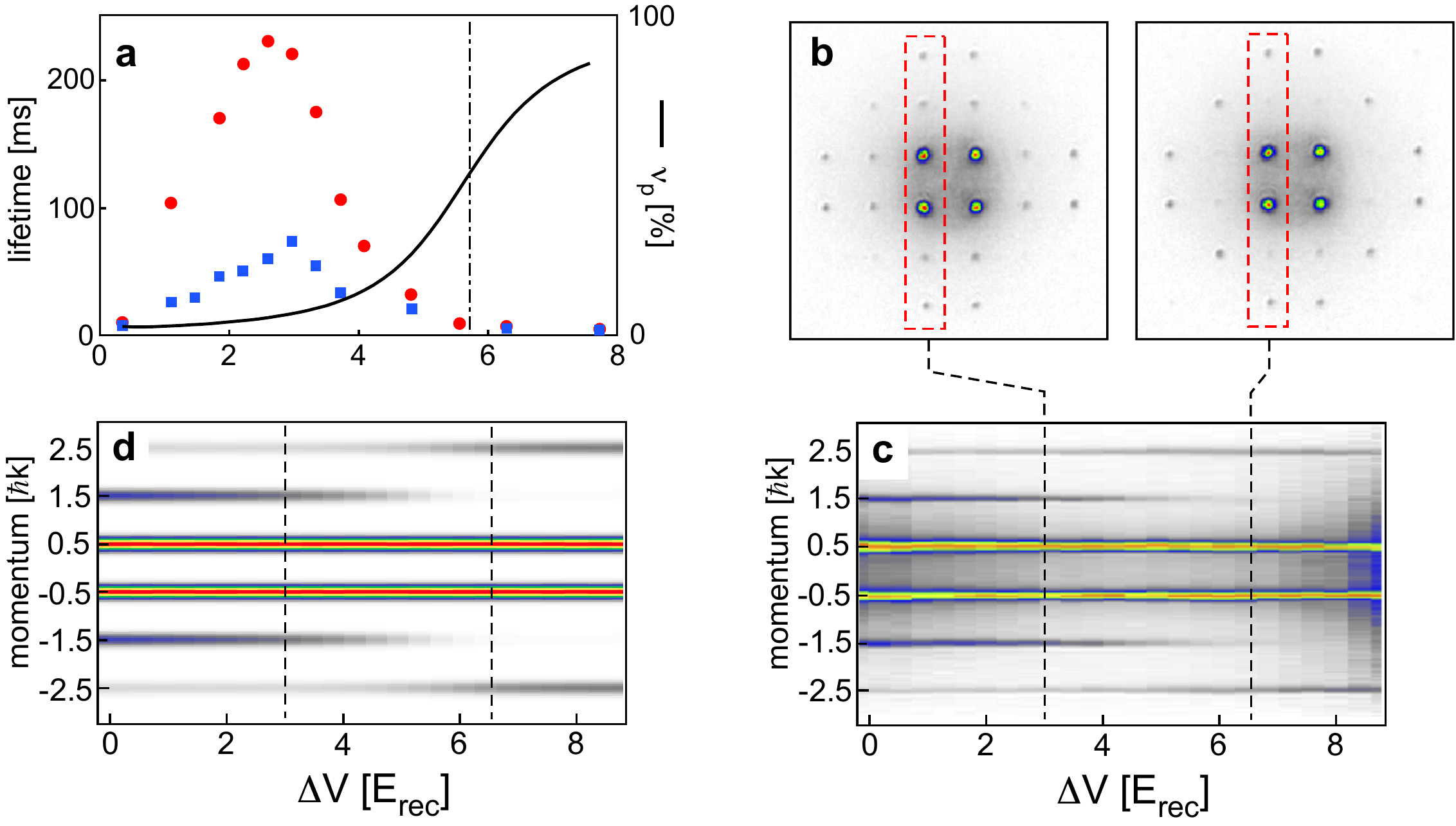}
\caption{\label{Fig.A2} (a) The data points show the lifetime of the atoms (red disks) and that of the condensed fraction (blue squares) in the second band versus $\Delta V$ for $V_0 = 6.0\, E_{\textrm{rec}}$. The solid line (derived from a band calculation) shows the relative $p$-orbital population $\nu_{p}$ versus $\Delta V$. The vertical dashed-dotted line indicates $\Delta V = V_{sp} = 5.7 \, E_{\textrm{rec}}$, where $\nu_{p}=\nu_{s} = 1/2$. (b) Two observations of momentum spectra recorded for small ($3.0\,E_{\textrm{rec}}$) and large $\Delta V$ ($6.5\,E_{\textrm{rec}}$). The red dashed rectangle identifies selected Bragg-maxima analyzed in more detail in (c) (observations) and (d) (theory) for varying values of $\Delta V$.}
\end{figure}
 
\begin{figure}
\includegraphics[width=160mm]{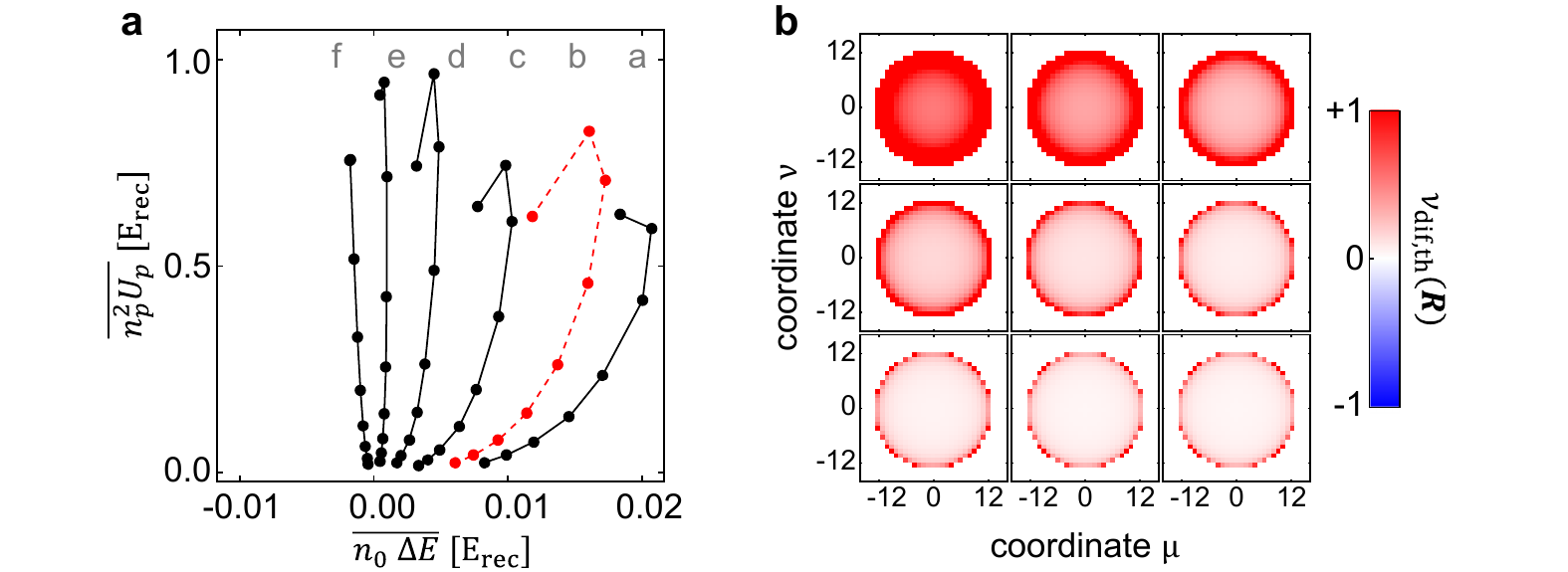}
\caption{\label{Fig.A3} (a) Map of the values of $\overline{n_p^2\,U_p}$ and $\overline{n_0 \,\Delta E}$ for the data points in the graphs (a)-(f) in Fig.~\ref{Fig.3} of the main text. The kinks in the paths arise due to collisional losses setting in for large values of $\nu_p$. (b) For the case of the red dashed path (b) corresponding to Fig.~\ref{Fig.3} (b), the local value of $\nu_{\text{dif,th}}(R)$ is plotted across the lattice. The lattice is parametrized by the Bravais vectors $R=\mu\,d_{+} + \nu\, d_{-}$.}
\end{figure}

The lifetime of the atoms in the second band (Fig.~\ref{Fig.A2} (a) red disks) is measured as follows: after forming the condensate at $\Delta V = 3 E_{\textrm{rec}}$, $\Delta V$ is tuned in 3 ms (sufficiently slow to permit tunneling) to the desired value. After a variable duration the population of the second band is determined (see V.). The wings of the decaying populations are fitted by exponentials with the $1/e$-times plotted as the red disks in Fig.~\ref{Fig.A2} (a). The lifetimes of the condensed fraction (blue squares in Fig.~\ref{Fig.A2} (a)) are obtained by an analog procedure, however counting the number of atoms in the lowest order Bragg peaks of a momentum spectrum.
\\ \\
\textbf{A VII. Calibrating the lattice distortion $\epsilon_y$.}
Changes of $\Delta E$ are experimentally implemented by adjustment of the lattice distortion parameter $\epsilon_y$ using polarization optics. Calibration of $\Delta E(\epsilon_y)$ proceeds as follows: the linear polarization of the incoming beam in the $y$-branch of the lattice potential is rotated such that $\langle n_+ \rangle \approx \langle n_- \rangle$, which corresponds to $\Delta E(\epsilon_y=1) = 0$. Arbitrary values of $\Delta E$ are adjusted by rotating the polarization away from this position by precisely quantified amounts and determining the corresponding values of $\Delta E$ via a band calculation for the potential in Eq.~(\ref{potential_A}). The estimated error in the determination of $\epsilon_y$ is $\Delta \epsilon_y \approx 2.5 \cdot 10^{-3}$.
\\ \\
\textbf{A VIII. Determination of $\overline{n_p^2 \, U_p}, \overline{n_0 \, \Delta E}$ and $\overline{\nu}_\text{dif,th}$.} We account for the isotropic harmonic potential with approximately 40 Hz trap frequency superimposed upon the two dimensional lattice of Eq.~(\ref{potential_A}) with $\eta = \epsilon_{x} = \epsilon_{y}=1$ and tubular lattice sites extending along the $z$-direction. The total number of condensed atoms, determined by fitting Gaussians to all visible Bragg peaks of a momentum spectrum and counting the atoms, is $N\approx 1.5 \cdot 10^4$ before band relaxation sets in. Assuming a Thomas-Fermi density distribution along the weakly confined $z$-direction, we calculate the local number of atoms per unit cell $n_{0,R}$ at the Bravais lattice site $R$, and hence the local number of particles in the $s$-orbitals and $p$-orbitals, $n_{s,R}=\nu_{s}\,n_{0,R}$ and $n_{p,R}=\nu_p\, n_{0,R}$, respectively. In the center of the lattice $n_{0,R} \approx 50$. In order to determine the local interaction energies per particle in the $p$-orbitals and $s$-orbitals at site $R$, $U_{p,R}$ and $U_{s,R}$, the local wells at the $\mathcal{A}$- and $\mathcal{B}$-sites are approximated to 6th order and the corresponding wave-functions of the local $p$-orbitals and $s$-orbitals are calculated. With this input, we determine the local energies $n_{p,R}^2 \,U_{p,R}$ and $n_{0,R}  \,\Delta E$. By applying Eq.~2 of the main text, the corresponding local value of $\theta$ and hence $\nu_\text{dif,th} = \sin^2(\theta)-\cos^2(\theta)$ is obtained. Averaging over all positions in the lattice yields $\overline{n_p^2 U_p}, \overline{n_0 \Delta E}$, used in Fig.~\ref{Fig.A3} (a) and $\overline{\nu}_\text{dif,th}$ plotted in Fig.~\ref{Fig.3} of the main text.
\\ \\
\textbf{A IX. Measurement of $\langle \nu_{\text{dif}}\rangle$ and comparison to theory $\overline{\nu}_\text{dif,th}$}. In order to obtain Fig.~\ref{Fig.3} (a)-(f) of the main text, in Eq.~(\ref{potential_A}) $\epsilon_y$ is adjusted to the six different values $1-\epsilon_y \in \{ 0.019, 0.014, 0.009, 0.004, 0.001, -0.001 \}$ and $\beta$ is varied for each setting. Momentum spectra are recorded and the number of atoms $n_{\pm}$ in the two zero order Bragg peaks corresponding to $X_{\pm}$ are counted. A data point in Fig.~\ref{Fig.3} (a)-(f) represents an average over 8 measurements. We thus obtain $\langle \nu_{\text{dif}}\rangle \equiv (\langle n_{+} \rangle-\langle n_{-} \rangle) / (\langle n_{+} \rangle + \langle n_{-} \rangle)$ for different combinations of the parameters $\beta$ and $\epsilon_{y}$. From $\beta$ one obtains $\Delta V$ (see Sec.A I) and thus $\nu_p$ (see Sec.A VI). The value $\Delta E$ is determined from a band calculation for the potential in Eq.~(\ref{potential_A}) (see Sec.A VII). For each data point, $n_0$ and $n_p = \nu_p n_0$ are derived at the time of the measurement, thus accounting for band relaxation loss. With this the energies $\overline{n_p^2\,U_p}, \overline{n_0 \,\Delta E}$ and the occupation difference $\overline{\nu}_\text{dif,th}$ are obtained (see Sec.A VIII). The data points in each of the graphs (a)-(f) in Fig.~\ref{Fig.3} of the main text correspond to a path in the plane spanned by the energies $\overline{n_p^2\,U_p}$ and $\overline{n_0 \,\Delta E}$. These paths are identified here in Fig.~\ref{Fig.A3} (a). For the case of the red dashed path corresponding to Fig.~\ref{Fig.3} (b), the local value of $\nu_{\text{dif,th}}$ is plotted across the lattice in Fig.~\ref{Fig.A3} (b). This confirms that, particularly for the data points in the center of the path, finite size effects are in fact substantial and have to be accounted for.
\\ \\
\textbf{A X. Mean field predictions of $\overline{\nu}_{\text{dif,th}}$ for real superposition state and incoherent mixture}. We have also minimized the total energy for the real coherent superposition $\sin(\theta) \, |\psi_{+}\rangle \pm \cos(\theta) \, |\psi_{-}\rangle$ (scenario \textbf{C}) and for the incoherent mixture of spatially superimposed condensates $\sin^2(\theta) \, |\psi_{+}\rangle \langle \psi_{+}| + \cos^2(\theta) \, |\psi_{-}\rangle \langle \psi_{-}|$ (scenario \textbf{D}) in order to calculate the local mixing angle $\theta$ as a function of $\Delta E$ and $\nu_p$ and hence to determine $\nu_\text{dif,th} = \sin^2(\theta)-\cos^2(\theta)$. For scenario \textbf{C} we obtain the simple result that $\nu_{\text{dif,th}}$ is constrained to the values $\pm 1$ depending on the sign of $\Delta E$ with no dependence upon $\nu_p$. For scenario \textbf{D}, values of $\nu_{\text{dif,th}}$ deviating from $\pm 1$ require $\nu_{p} > 2/3$ and satisfy
\begin{eqnarray}
\label{mixture}
\nu_{\text{dif,th}} =  \frac{3 \, \Delta E}{(\nu_{p}^2-4\,(1-\nu_{p})^2)\,n_{0}\,U_{p}} \,.
\end{eqnarray}
An average over all plaquettes of the lattice as described in Sec.A VIII leads to the corresponding value of $\overline{\nu}_{\text{dif,th}}$, which yields the (blue) dashed traces in Fig.~\ref{Fig.3} of the main text.

\section{Acknowledgements}
This work was partially supported by DFG-He2334/14-1, DFG-SFB 925, DFG-GrK1355, and the Netherlands Organization for Scientific Research (NWO). We are grateful to L. Mathey and W. Vincent Liu for useful discussions.


\section{References}

\begin{thebibliography}{28}

\bibitem{Tok:00}
Tokura Y and Nagaosa N 2000 \textit{Science} {\bf 288} 462-8

\bibitem{Mae:94}
Maeno Y, Hashimoto H, Yoshida K, Nishizaki S, Fujita T, Bednorz J G, Lichtenberg F
1994 \textit{Nature} {\bf 372} 532-4

\bibitem{Bee:12}
Beenakker C W J 2013 \textit{Annu. Rev. Con. Mat. Phys.} {\bf 4} 113-36

\bibitem{Lew:07}
Lewenstein M, \textit{et al} 2007 \textit{Adv. Phys.} {\bf 56} 243-379

\bibitem{Blo:08}
Bloch I, Dalibard J and Zwerger W 2008 \textit{Rev. Mod. Phys.} {\bf 80} 885-964

\bibitem{Fey:72}
Feynman R P 1972 Statistical Mechanics: A Set of Lectures. Addison-Wesley Publishing Company

\bibitem{Wu:09}
Wu C 2009 \textit{Mod. Phys. Lett.} B {\bf 23}, 1-24

\bibitem{Lin:09}
Lin Y-J, Compton R L, JimŽnez-Garc'a K , Porto J. V. and Spielman I B 2009 \textit{Nature} {\bf 462} 628-32

\bibitem{Lin:11}
Lin Y-J, JimŽnez-Garc'a K, Porto J V  and Spielman I B 2011 \textit{Nature} {\bf 471}, 83-6

\bibitem{Dal:11}
Dalibard J, Gerbier F. Juzeli\=unas G and \"{O}hberg, P 2011 \textit{Rev. Mod. Phys.} {\bf 83} 1523-43

\bibitem{Aid:11}
Aidelsburger M, Atala M, Nascimb\'{e}ne S, Trotzky S, Chen Y-A and Bloch I  2011 \textit{Phys. Rev. Lett.}{\bf 107} 255301 

\bibitem{Eck:05}
Eckardt A, Weiss C and Holthaus M 2005 \textit{Phys. Rev. Lett.}{\bf 95} 260404

\bibitem{Zen:09}
Zenesini A, Lignier H, Ciampini D, Morsch O and Arimondo E 2009 \textit{Phys. Rev. Lett.}{\bf 102} 100403

\bibitem{Str:11}
Struck J, \"{O}lschl\"{a}ger C, Le Targat R, Soltan-Panahi P, Eckardt A, Lewenstein M, Windpassinger P and Sengstock K 2001 \textit{Science} {\bf 333} 996-99 

\bibitem{Isa:05} 
Isacsson, A and Girvin S 2005 \textit{Phys. Rev. A} {\bf 72} 053604

\bibitem{Liu:06} 
Liu W V and Wu C 2006 \textit{Phys. Rev. A} {\bf 74} 013607 

\bibitem{Wir:11}
Wirth G, \"{O}lschl\"{a}ger M and Hemmerich A. 2001 \textit{Nature Physics} {\bf 7} 147-53

\bibitem{Hem:91}
Hemmerich A, Schropp D and H\"{a}nsch T W 1991 \textit{Phys. Rev. A} {\bf 44} 1910-21

\bibitem{Oel:11}
\"{O}lschl\"{a}ger M, Wirth G and Hemmerich A 2011 \textit{Phys. Rev. Lett.}{\bf 106} 015302 

\bibitem{Tie:10} 
The first equality has been derived with contributions by O. Tieleman. 

\bibitem{Cai:11}
Cai Z and Wu C 2011 \textit{Phys. Rev. A} {\bf 84} 033635 

\bibitem{Liu:12} 
Liu B, Yu X-L and Liu W-M 2012 arXiv:1211.2595v3

\bibitem{Cai:12}
Cai Z, Duan L-M and Wu C 2012 \textit{Phys. Rev. A} {\bf 86} 051601(R) 

\bibitem{She:10}
Sherson J F, Weitenberg C, Endres M, Cheneau M, Bloch I and Kuhr S 2010 \textit{Nature} {\bf 467} 68-72

\bibitem{Gre:10}
Bakr W S, Peng A, Tai M E, Ma R, Simon J, Gillen J I, F\"{o}šlling S, Pollet L and Greiner M 2010 \textit{Science} {\bf 329} 547-50

\bibitem{Li:11}
Li X, Zhao E and Liu W V 2011 \textit{Phys. Rev. A} {\bf 83} 063626

\bibitem{Pin:12}
Pinheiro F, Martikainen J-P and Larson J 2012 \textit{Phys. Rev. A} {\bf 85} 033638

\bibitem{Mar:12}
Martikainen J-P and Larson J 2012 \textit{Phys. Rev. A} {\bf 86} 023611

\bibitem{Oel:12}
\"{O}lschl\"{a}ger M, Wirth G, Kock T and Hemmerich A 2012 \textit{Phys. Rev. Lett.} {\bf 108} 075302 

\bibitem{Sun:12}
Sun K, Liu W V, Hemmerich A and Das Sarma S 2012 \textit{Nature Physics} {\bf 8} 67-70 

\end{thebibliography}
\end{document}